\documentstyle{mn}

\newcommand{\pg}{\mbox{PG\,0859+415}}

\newcommand{\ha}{\hbox{$\hbox{H}\alpha$}}

\newcommand{\hei}{\hbox{$\hbox{He\,{\sc i}\,$\lambda$4471\,\AA}$}}
\newcommand{\heired}{\hbox{$\hbox{He\,{\sc i}\,$\lambda$5876\,\AA}$}}

\newcommand{\kmsec}{\,\mbox{$\mbox{km}\,\mbox{s}^{-1}$}}

\title[The cataclysmic variable PG\,0859+415]
      {An accretion model for the eclipsing cataclysmic variable 
       PG\,0859+415}
\author[M.\,D.\ Still]
       {M.\,D.\ Still$^{1,2}$ \\
       $^1$Physics and Astronomy, University of St.\ Andrews, North Haugh,
       St.\ Andrews, Fife KY16\,9SS (mds1@st-and.ac.uk) \\
       $^2$Astronomy Centre, School of Mathematical and 
       Physical Sciences, University of Sussex, Falmer, 
       Brighton BN1\,9QH} 
\date{Accepted 1996 May 20.
      Received 1996 April 25;
      in original form 1996 January 8}

\begin{document}

\maketitle

\begin{abstract} 

The emission lines found in the majority of cataclysmic variables are
generally used as tracers of accretion flows which dominate the light at
optical wavelengths. It has been suggested from previous observations that
the eclipsing nova-like variable PG\,0859+415 shows individualistic
orbital characteristics which are inconsistent with canonical models of
stellar accretion. We present spectrophotometry of this star which suggest
that the standard picture is not in conflict with observation. We provide
evidence that the shallow optical eclipses are of an extended bright spot
rather than the accretion disc, and that the low-excitation lines are
dominated by a transient absorption component, perhaps a result of
the bright spot or accretion stream eclipsing the disc. We argue that this
object may be another nova-like variable displaying long-term or cyclic 
variations in mass-transfer rate from the secondary star. 

\end{abstract}
 

\section{Introduction}

\pg\ is a 14th magnitude, eclipsing nova-like variable with an orbital
period of 3.7\,h (Grauer et~al. 1994). As such, it is a member of the
cataclysmic variable family, consisting of a red dwarf (the secondary star)
transferring mass onto a white dwarf companion (the primary star) via Roche
lobe overflow. Since there is no observational evidence for a
significant magnetic field about the compact object, accretion is assumed
to occur via a disc, which must be stable to variations in mass transfer
rate in order to explain the lack of observed dwarf nova outbursts. A
recent observational review of the nova-likes is provided by Dhillon
(1996). 

The majority of cataclysmic variables are emission line objects, which
are used as tracers of the accretion flow that dominates the spectrum at
optical wavelengths. In dwarf novae, emission features are
superficially understood to be composites of lines from the accretion
disc and the bright spot, where the gas stream from the secondary star
impacts the outer rim of the disc (e.g.\ Marsh et~al.\ 1990). Similar
observations of the more energetic nova-likes, however, provide a
complex picture which is proving difficult to unravel. Optical lines
appear to be composites of both emission and absorption line spectra
from a host of possible sources, such as the disc (Rutten et~al. 1994),
a disc wind (Honeycutt, Schlegel \& Kaitchuck 1986), the bright spot
(Still, Dhillon \& Jones 1995), the gas stream overflowing the accretion
disc (Hellier \& Robinson 1994), and the irradiated inner face of the
secondary star (Beuermann \& Thomas 1990). Given such a range of possible
contributions to optical spectra, it is unsurprising that nova-like
emission shows such diversity in properties and that the nova-like
sub-classification scheme is constantly being questioned. 

The blue excess object \pg\ was confirmed as a cataclysmic variable
during the optical campaign of Grauer et~al.\ (1994; hereafter G94).
Broad-band photometry showed rapid flickering and eclipses every 220\,m,
which was subsequently interpreted as the orbital period. Time-resolved
spectroscopy reveals a high-excitation emission line spectrum with the
higher Balmer series containing shallow absorption wings, presumably
from the thick inner disc, and which display velocity modulation on the
orbital period. The long-term constancy of light curves combined with
the emission profiles suggest a nova-like classification. 

Their data provided G94 with a number of
provocative ideas. Firstly, photometry shows that eclipses are preceded
by a broad hump, consistent with behaviour expected if the bright spot
contributes to the optical flux. However, the depth of eclipse never
exceeded the pre-hump photometric level -- suggesting the possibility
that the eclipse is of the bright spot, rather than the disc. Secondly,
study of Balmer line profiles reveal a prominent narrow emission source
with orbital-dependent velocities consistent with localized gas on the
near side of the disc when observed at secondary star superior conjunction.
Such a source is not predicted by models of disc accretion. 

In this paper we present spectrophotometry of \pg, consistent with the
data obtained by G94, and suggest an acceptable emission model which is
a composite of secondary star irradiation, gas stream absorption and emission
from a wind and/or an accretion disc. We find further evidence that the
eclipsed source is the bright spot rather than the disc, providing an
opportunity to study variations in mass transfer rate from the secondary star
via changes in spot properties and disc size. 

\section{Observations}

\pg\ was observed on the nights beginning 1995 Feb 16-18 with the 2.5\,m
Isaac Newton Telescope on La Palma. Exposure times began at 140\,s on
Feb 16 but were increased over the night to optimize signal and orbital
resolution. Dwell times of 400\,s were decided upon and used for the
rest of the run. Dead time for the dumping of data was generally 70\,s
between frames. The detector was a Tektronix CCD, windowed to
300\,$\times$\,1124 pixels and mounted on the Intermediate Dispersion
Spectrograph with a grating of 1200~lines\,mm$^{-1}$.  This
configuration gave a wavelength range of approximately
$\lambda\lambda$5811--6671\,\AA\ at 1.7\,\AA\ resolution. A journal of
observations is given in Table~1. 
\begin{table}
\centering
\begin{minipage}{84mm}
\caption{Journal of observations. E is the cycle number plus binary 
phase with respect to the ephemeris derived in Sec.~3.1.} 
\begin{center}
\begin{tabular}{cccccc}
\multicolumn{1}{c}{Date} & \multicolumn{1}{c}{Start} & 
\multicolumn{1}{c}{End} & \multicolumn{1}{c}{Start} & 
\multicolumn{1}{c}{End} & \multicolumn{1}{c}{No. of} \\
\multicolumn{1}{c}{(1995 Feb)} & \multicolumn{2}{c}{(UT)} & 
\multicolumn{2}{c}{(E $-$ 12\,000)} & \multicolumn{1}{c}{spectra} \\ \hline
16/17 & 23.66 & 3.90 & 326.43 & 327.58 & 46 \\
17/18 & 22.77 & 3.94 & 332.73 & 334.14 & 40 \\
18/19 & 22.84 & 3.98 & 339.29 & 340.69 & 35 \\
\end{tabular}
\end{center}
\end{minipage}
\end{table}
A nearby comparison star was placed on the slit and several photometric
spectra were obtained through a wide slit for the purpose of flux
corrections. CuAr arc exposures were taken every 30--40\,m in order to
calibrate the wavelength scale and instrumental flexure. The flux
standard, Feige~66 (Massey et~al.\ 1988), was observed with a wide slit
to correct for instrumental response and convert the spectra to an
absolute scale. 
 
Medium-scale sensitivity variations were removed with a balance frame
prepared from tungsten lamp and sky flat-fields. Fourth-order polynomial
fits to the sky were subtracted and raw spectra were then extracted
using the optimal algorithm of Horne (1986). Arc spectra for \pg, the
comparison star and the flux standard were extracted from appropriate
spatial locations on the arc frames. Fifth-order polynomial fits were
made to the arc lines (rms scatter~$<$~0.05\AA) and the wavelength scale
for each spectrum was obtained by interpolating between two consecutive
arc spectra. A comparison of the tabulated absolute flux values with a
spline fit to the continuum of the flux standard provided a correction
for large-scale instrumental response. Slit-loss corrections were
achieved by dividing the \pg\ spectra by a constant found from summing
the flux under the continuum of the corresponding comparison star observation
and multiplying by a constant derived similarly from the average of the
comparison star wide-slit exposures. A wavelength-dependent correction proved
to be impossible due to the low photon counts in individual comparison star
spectra. 
 
\section{Results}

\subsection{Eclipse timings}

\begin{figure}
\begin{picture}(100,0)(10,20)
\put(0,0){\includegraphics{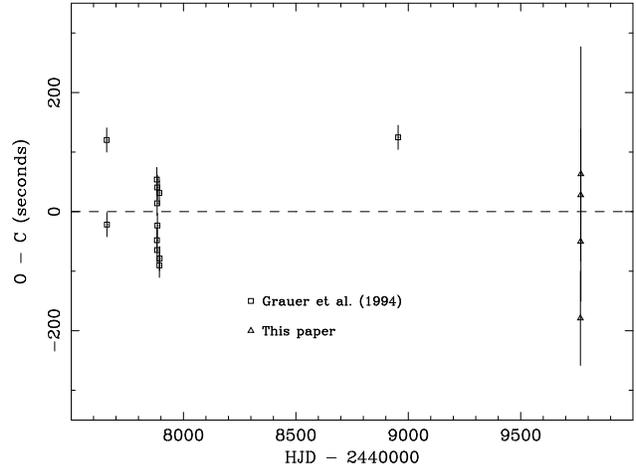}}
\noindent
\end{picture}
\vspace{62mm}
\caption{$O-C$ diagram for the times of mid-eclipse of \pg. The dashed
line defines the linear ephemeris refined in this paper, the square
symbols denote the photometric timings of G94 and the triangles are the
spectrophotometric timings from the current data.}
\end{figure}

Four eclipses were covered by these observations, providing an
opportunity to test and refine the orbital ephemeris measured by G94.
Eclipse profiles, which were produced by summing over line-free
wavelengths of the continuum, were fitted with parabolae (details of
light curve production may be found in Sec.~3.3). The minima of these,  
presented in Table~2, were combined with the photometric timings
of G94 and fit with a linear function, giving an apparently improved
orbital ephemeris of: 
\[ 
T_0 = {\mbox{HJD}}\,2447881.8584(7) + 0.15281249(2)\,E
\]
where $T_0$ is the time of mid-eclipse, and $E$ is the subsequent
orbital cycle number. 

\begin{table}
\centering
\begin{minipage}{84mm}
\caption{$O-C$ timings for mid-eclipses of \pg, relative to the new 
ephemeris.} 
\begin{center}
\begin{tabular}{clcr}
\multicolumn{1}{c}{Date} & \multicolumn{1}{c}{HJD} & 
\multicolumn{1}{c}{Cycle} & \multicolumn{1}{c}{$O-C$} \\
\multicolumn{1}{c}{(1995 Feb)} & \multicolumn{1}{c}{} & 
\multicolumn{1}{c}{(E)} & \multicolumn{1}{c}{(s)} \\ \hline
17 & 2\,449\,765.5760(9)  & 12\,327 & $-$179(79)\hspace{.5em} \\
17 & 2\,449\,766.4943(3)  & 12\,333 & $-$50(24)\hspace{.5em} \\
18 & 2\,449\,766.6481(13) & 12\,334 & \hspace{1em}28(110) \\
19 & 2\,449\,767.5653(24) & 12\,340 & \hspace{1em}63(213) \\
\end{tabular}
\end{center}
\end{minipage}
\end{table}

The residuals of this fit are presented in the $O-C$ diagram of Fig.~1.
It is unclear whether the presence of outliers is systematic or
statistical, but their significance is discussed in Sec.~4. 

\newpage

\subsection{Average spectrum}

\begin{figure}
\begin{picture}(100,0)(10,20)
\put(0,0){\includegraphics{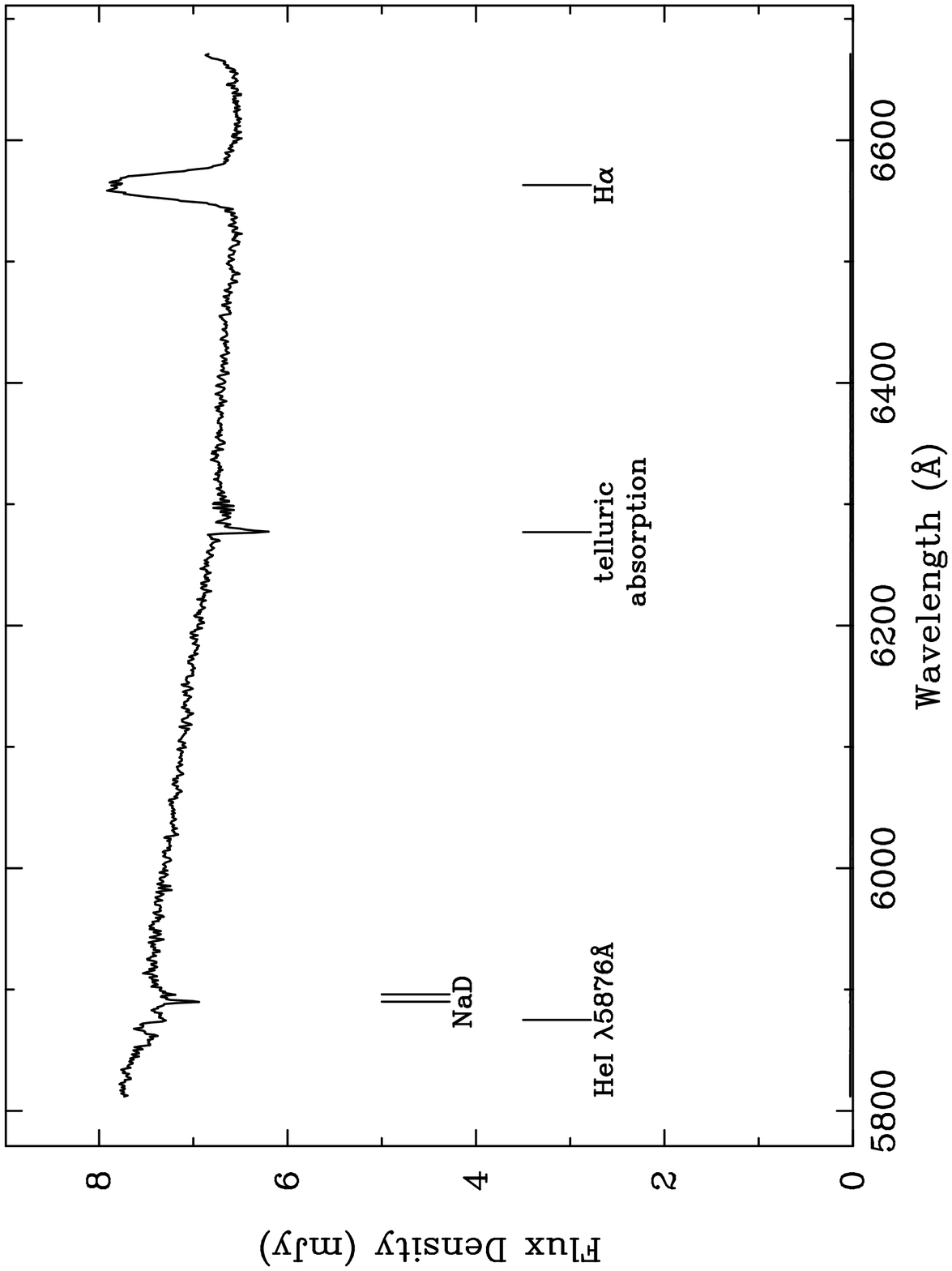}}
\noindent
\end{picture}
\vspace{62mm}
\caption{The average spectrum of \pg.}
\end{figure}

The average of all spectra of \pg\ is presented in Fig.~2. Given that the
radial velocity of either stellar component remains unknown, spectra
have not been re-binned to remove orbital line smearing.  We find a
strong blue continuum, superposed with weak emission lines of \ha\ and
\heired. \heired\ displays an emission core with shallow absorption
wings and the red wing is blended with interstellar absorption. The
strong continuum and He{\sc i} profile suggest that a relatively large
fraction of the disc is projected on the sky, compared to the deeply
eclipsing members of the class (e.g.\ UX~UMa: Rutten et~al.\ 1994). The
\ha\ profile is single peaked, which is at odds with profiles
of accretion disc emission, but this is not unexpected. 

\subsection{Spectrophotometry}

\begin{figure}
\begin{picture}(100,0)(10,20)
\put(0,0){\includegraphics{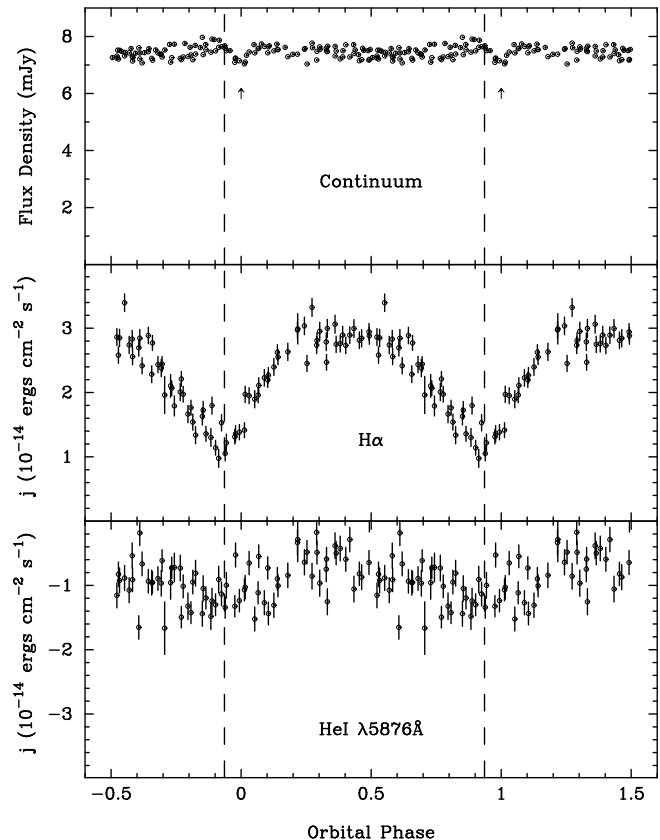}}
\noindent
\end{picture}
\vspace{114mm}
\caption{Orbital light curves for the continuum, \ha\ and \heired\
from \pg. Each curve is repeated over a second orbital cycle and 
orbital phase~0.0 corresponds to mid-eclipse and is marked by an arrow; 
however,
the dashed lines, which are measured from Doppler images, are believed to be
close to the true phase of stellar conjunction (Sec.~3.4).}
\end{figure}

A continuum light curve was produced by summing each exposure over the
line-free wavelengths $\lambda\lambda$\,5929--6257\,\AA,
$\lambda\lambda$\,6358--6506\,\AA\ and
$\lambda\lambda$\,6625--6646\,\AA. This is presented in Fig.~3. Both the
pre-eclipse hump and shallow eclipse are found in this light curve,
although they appear less obvious than in the magnitude plots presented
by G94 because of the linear scale, the range chosen to present the data, and
nightly variations. 
Consistent with G94's findings, the eclipse depth does not significantly
exceed the pre-hump continuum level. 

Line light curves were produced by subtracting a spline fit to the
line-free continuum from the data, and summing over the profiles. There
is no significant evidence for eclipse above the noise limits of either
feature.  The \ha\ curve is similar to those of nova-likes whose lines
are dominated by a component from the irradiated inner face of the
secondary star (e.g.\ Dhillon, Jones \& Marsh 1994). Given that stellar
conjunction may not necessarily coincide with mid-eclipse (Sec~3.4), it
is still true that maximum emission occurs at approximately the superior
conjunction of the secondary star when, the inner, irradiated face of the 
star
is most visible. The same behaviour may be found in the \heired\ line,
although the flux contribution is dominated by the absorption wings,
most probably originating in the inner accretion disc. 

\begin{figure} 
\begin{picture}(100,0)(10,20)
\put(0,0){\includegraphics{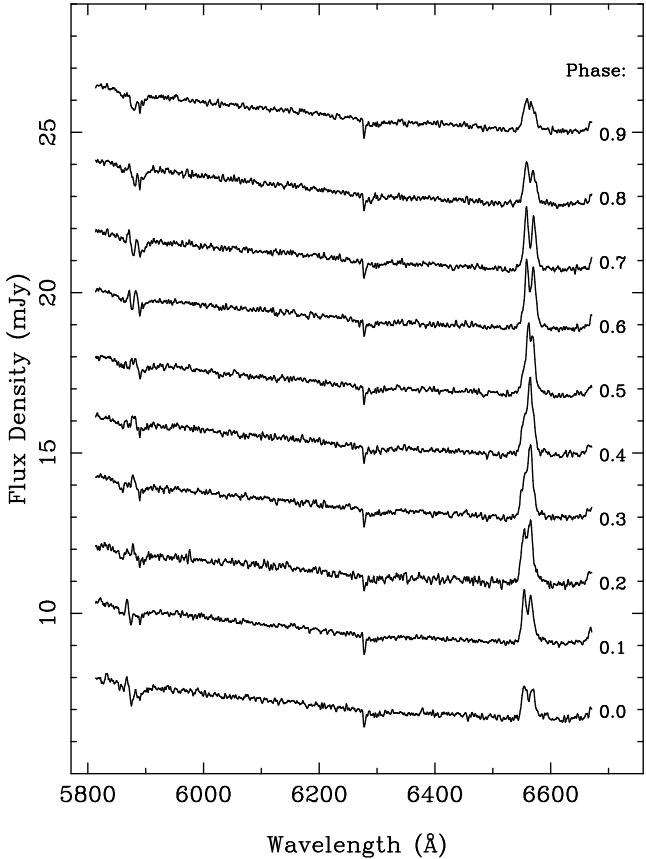}}
\noindent
\end{picture}
\vspace{116mm}
\caption{Orbital phase-binned spectra of \pg, stacked with offsets
which are multiples of 4\,mJy.} 
\end{figure}

If the secondary star is the major component of line emission, it will be
evident in the orbitally-resolved line profiles. The data were averaged
into phase bins, each covering 1/10 of the orbital cycle. Multiple offsets
were applied to each and the resulting stack is plotted in Fig.~4. There
appears to be at least two components to the \ha\ line, suggesting that
a line model containing only secondary star emission is inadequate. The
profile consists of two anti-phased peaks, one consistent with an origin
from the secondary star, and the other consequently associated with the
primary star. This second peak is the feature identified by G94 as emission
from a source at the front of the accretion disc, observed at superior
conjunction of the secondary star. It should be noted however, that the profile
can also be interpreted as one, or a composite of emission features, with
a variable absorption component contaminating the core. Despite the small
signal, the \hei\ feature displays the same behaviour superposed over the
broad absorption wings. 

\subsection{Line profile fitting}

The data sampling and spectral resolution allow a more detailed analysis of
the emission line profiles than the coarse phase bins of Fig.~4 allow. The
data were continuum subtracted in the usual way, re-binned onto a constant
velocity scale of 39\,\kmsec, and averaged into orbital phase bins each
covering 2\% of the orbit for \ha, and 5\% for \heired. The line profiles
are presented as functions of orbital phase in Fig.~5. The linear
grey-scale goes from white to black with increasing line intensity but has
different zero points for each image, 0--3\,mJy for \ha, and
$-$0.6--0.1\,mJy for \heired. The data are repeated over a second orbital
cycle and blank strips represent unfilled phase bins. 

We first consider the \ha\ feature, on which G94 base their analysis. The
profile is dominated by two anti-phased emission components.  The
component with smaller velocity amplitude is approximately consistent with
emission from the secondary star (Dhillon et~al.\ 1994). Consequently,
assuming a stellar mass ratio ($q$~=~$M_{\mbox{\sevensize 
s}}/M_{\mbox{\sevensize
p}}$) less than unity, the anti-phased emission is consistent with a
location on the back of the disc as viewed at the superior conjunction of the 
primary star, and as indicated by G94. Both emission components appear
strongest at phase~0.5, as expected if the inferred spatial location of
each source is correct. At orbital phases when the anti-phased emission
does not dominate the profile, the line generally appears double-peaked,
suggestive of a disc origin, or of absorption contamination from a local
source within the accretion flow contaminating the core (Still et~al.\
1995). 

Further insight into the line sources is often provided by image
reconstruction using Doppler tomography (Marsh \& Horne 1988). This
involves the mapping of velocity--phase data ($V$,$\phi$) to a
velocity--velocity field ($V_x$,$V_y$) via: 
\[
f(V,\phi) = \int_{-\infty}^\infty \int_{-\infty}^\infty I(V_x,V_y)~\times
\]
\begin{equation}
{\mbox{\hspace{6.6em}}}g(V - \gamma + V_x \cos{\phi} - V_y 
\sin{\phi})\,dV_x\,dV_y 
\end{equation}
where $f$ is the line intensity at velocity $V$ and orbital phase $\phi$,
$I$ is the emission distribution of the resulting map in velocity
coordinates ($V_x$,$V_y$), $\gamma$ is the systemic velocity and $g(V)$ is
the local line profile. Examples of similar image reconstructions may be
found in e.g.\ Marsh \& Horne (1990). Given that the profile of \ha\
possibly contains an absorption component, the preferred method is that of
Fourier-filtered back-projection (Horne 1992), thereby avoiding the positivity
constraint demanded on reconstructions by a maximum entropy approach. 

\begin{figure*} 
\begin{picture}(100,0)(10,20)
\put(0,0){\includegraphics{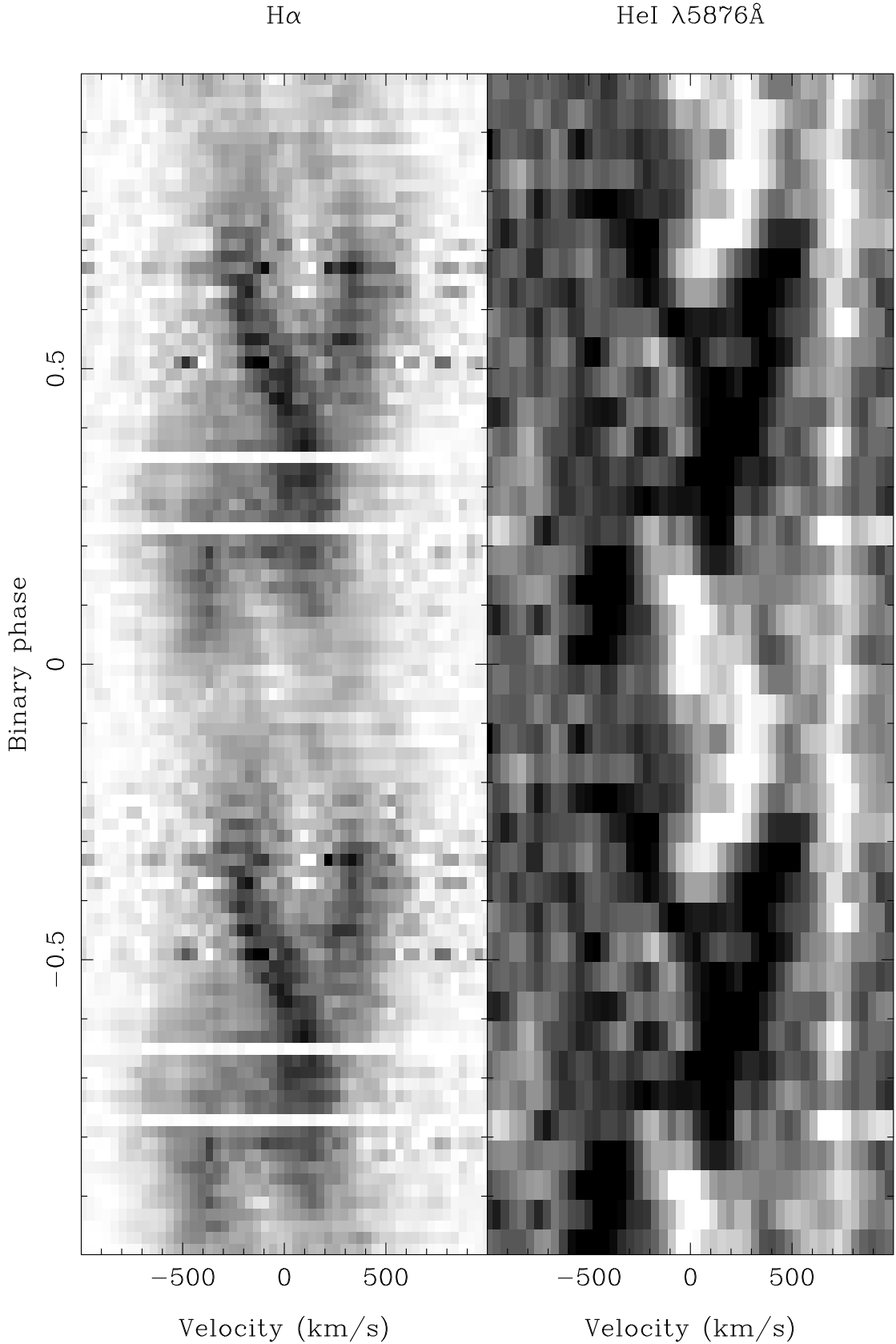}}
\noindent
\end{picture}
\vspace{227mm}
\caption{Trailed spectrograms of the \ha\ and \heired\ lines from \pg. The 
data has been repeated over a second orbital cycle and blank strips indicate
unfilled phase bins. The two images are on separate linear intensity 
scales.} 
\end{figure*}

\begin{figure*} 
\begin{picture}(100,0)(10,20)
\put(0,0){\includegraphics{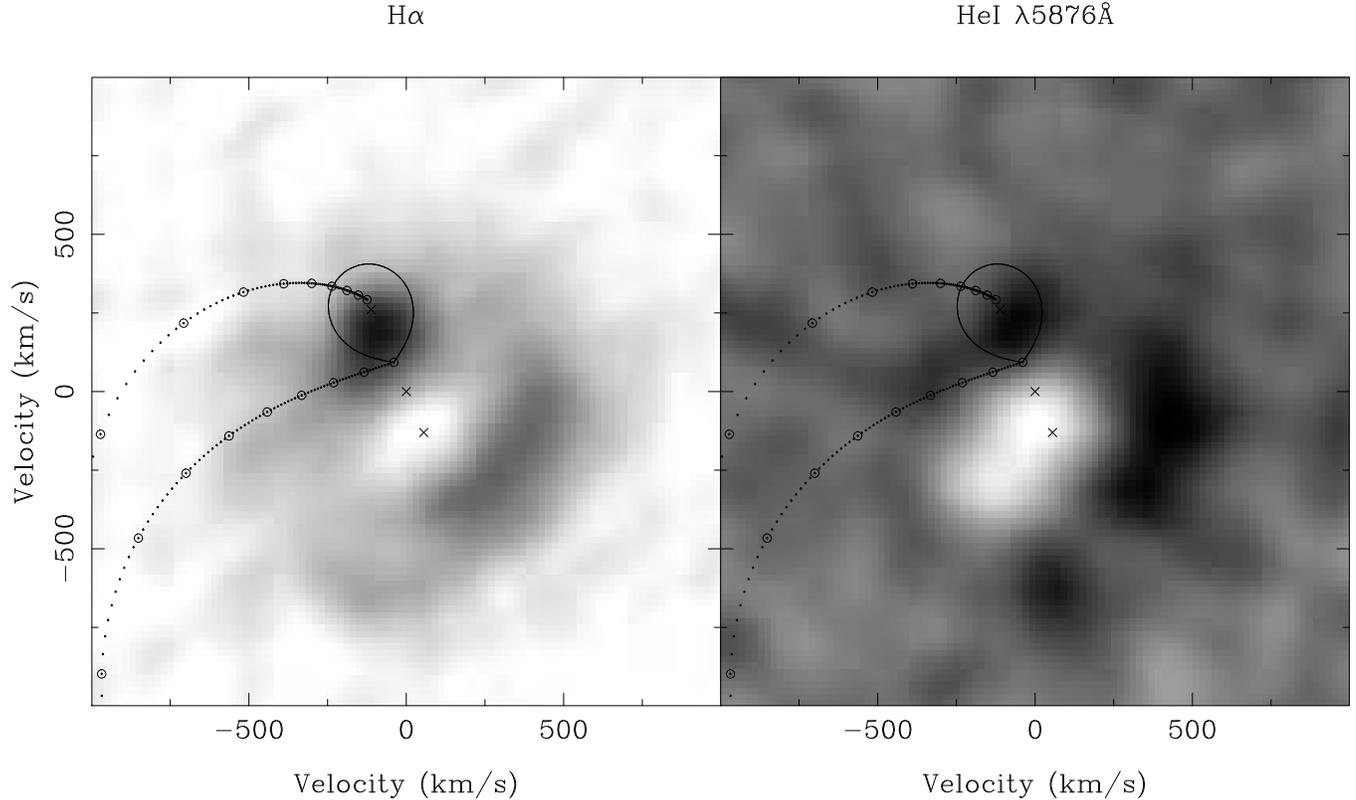}}
\noindent
\end{picture}
\vspace{110mm}
\caption{Doppler tomograms of the \ha\ and \heired\ feature from \pg. The 
intensity scales are linear but different for the two maps, where white 
to black illustrates increasing flux.} 
\end{figure*}

\begin{figure*} 
\begin{picture}(100,0)(10,20)
\put(0,0){\includegraphics{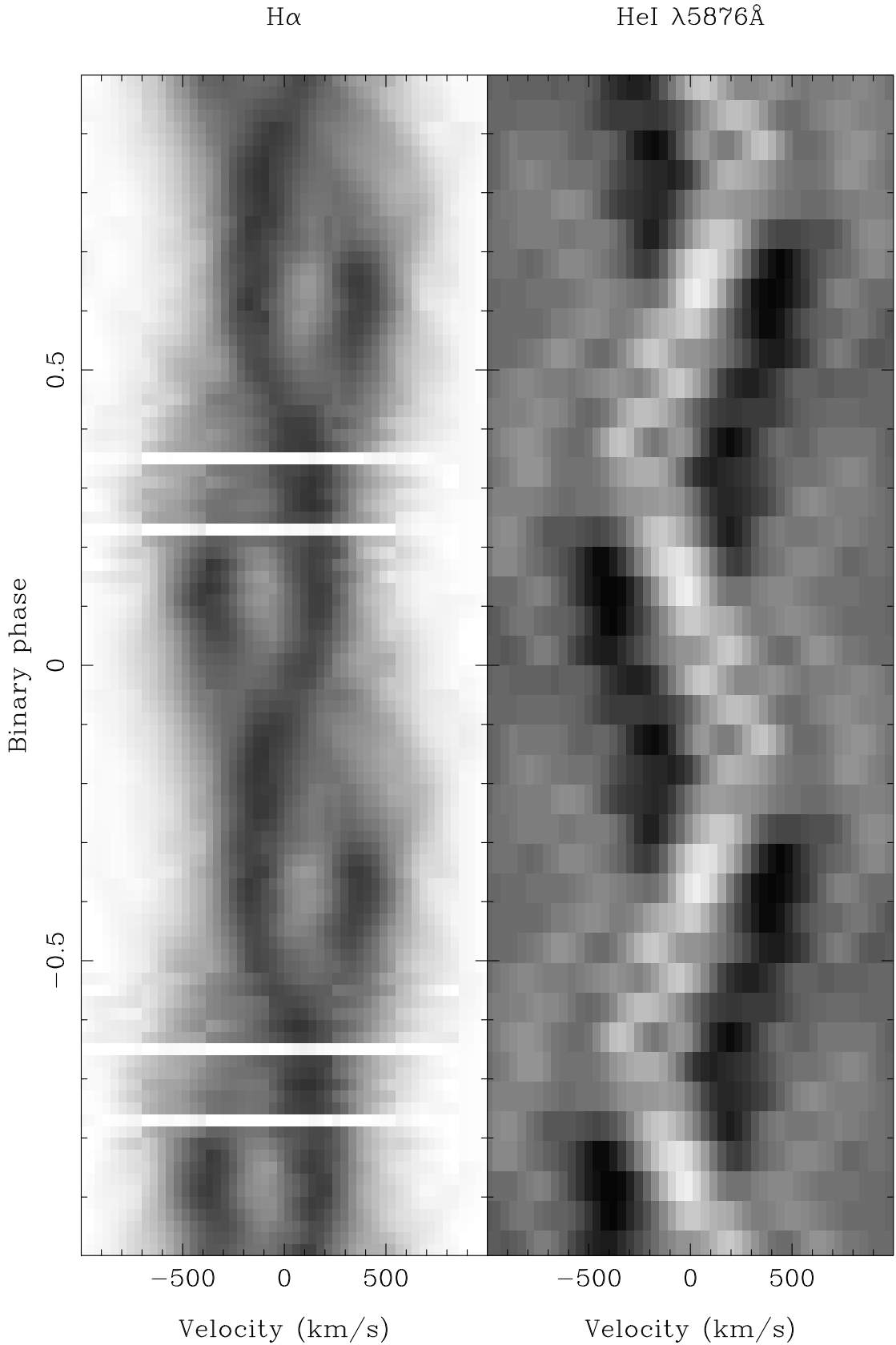}}
\noindent
\end{picture}
\vspace{224mm}
\caption{Model spectrograms of the \ha\ and \heired\ lines from \pg, 
computed from the tomograms of Fig.~6. The 
data have been repeated over a second orbital cycle and blank strips indicate
unfilled phase bins. The two images are on separate linear intensity 
scales.} 
\end{figure*}

In order to reduce smoothing in the map,
the data were convolved with a high-pass filter damped with a Gaussian 
cutoff to avoid high-frequency noise amplification. Back-projection results 
in the velocity map presented in Fig.~6. $V_x$ is the abscissa and $V_y$ 
the ordinate. Superposed over the image 
is an approximate binary configuration, where $K_{\mbox{\sevensize
s}}$~=~280\,\kmsec\ and $q$~=~0.5. The secondary Roche lobe is plotted,
with the trajectory of the accretion stream, originating at the inner
Lagrangian point, and the Keplerian velocity about the white dwarf along
the stream, originating from close to the secondary star's centre of mass. 
This second trajectory represents the velocity of a Keplerian accretion
disc at the spatial location of the stream (Marsh et~al.\ 1990). The three
crosses represent the centres of mass for the secondary star, binary and
primary star. An orbital solution for \pg\ is not known with any confidence and
the choice of binary parameters is discussed shortly. 

The features within this map are as expected from the trailed data. There 
are two major emission spots, one which we interpret as originating from 
the inner face of the the secondary star and the other, if we consider the 
velocity field to be that of a Keplerian accretion disc about the white 
dwarf, from a local source on the back of the accretion disc. A weaker 
ring is approximately centred on the primary star, suggestive of accretion 
disc emission. Fits to the line profiles, computed from the tomogram, are 
presented in Fig.~7. These are useful as a guide to the quality of the 
mapping procedure. Note, however, that pixel values in the map are not 
time-variable, and therefore the tomogram struggles to reconstruct 
the flux variations of individual line components within the data.

Clearly, the secondary star emission in the \ha\ tomogram is not centred over
the $V_x$~=~0 axis. Provided we are willing to accept the secondary star as the
origin for this feature, the most immediate explanation is that the
eclipse ephemeris does not measure times of stellar conjunction. The
consequent direction of coordinate rotation is consistent with the
hypothesis of G94 that optical eclipses are of the bright spot only. A
Gaussian fit to the secondary star peak indicates maximum emission at a
velocity of 175\,$\pm$\,4\,\kmsec, and 0.064\,$\pm$\,0.002 orbital cycles
before mid-eclipse. Since the secondary star emission is most likely to
originate on its inner face, the measured velocity may be taken as a lower
limit to the stars radial velocity semi-amplitude. 
The measured phase is an approximation
of white dwarf superior conjunction, which is illustrated by both the
orientation of the binary on the tomograms and the dashed line found in
Fig.~3. However, if irradiation of the secondary star is not symmetric about
the $L_1$-point, perhaps a consequence of shadowing by the bright spot or
accretion stream (Southwell et~al.\ 1995), then the conjunction phase will
be inaccurate by no more than a few percent of the orbital period. The
\ha\ data are not suitable for providing a measurement of the primary stars
radial velocity semi-amplitude since it remains unclear whether any of the emission is
distributed symmetrically about the white dwarf (Shafter 1985; Still
1996). Therefore the binary parameters adopted for Fig.~6 are a guess,
based on a mass ratio of $q$~=~0.5 and assuming that only the inner face
of the secondary star is irradiated. 

The emission spot at the back of the disc has been inferred from two
independent data sets, yet there appears no plausible reason for its
presence. A solution is provided from an inspection of the \heired\
profiles, which suggest that the spot is an artifact of core absorption. 
Fig.~5 illustrates that the two emission features found in the \ha\ line
are also observed in the He\,{\sc i} profile. Also present, however, is an
absorption component which dominates the line core and has variable
velocity and intensity over the orbital cycle, attaining maximum strength
between orbital phases~0.6 and 1.1 and maximum velocity at phase~0.9. This
provides a consistent, yet alternative, model for the line 
distribution: The first component is a broad
emission feature which is visible in Fig.~5 between $-$500\,\kmsec\ and
+500\,\kmsec; this is most likely from the accretion disc (Rutten et~al.\
1994), or an accretion disc wind (Dhillon, Marsh \& Jones 1991). The second
is an absorption line overlaying the first component, similar to the
feature observed in the deeply eclipsing nova-like RW~Tri (Still et~al.\
1995), and the third is narrow emission from the secondary star (Dhillon
et~al.\ 1994). The last component appears to be faint but still visible at
phase~0.9 where it is observed to cross the absorption feature. This is
plausible given the inclination inferred by the shallow eclipse, and the
superposition of this narrow component over the absorption provides
confidence that it is an individual emission component and not the residue of 
core absorption.

With hindsight, this line model is equally plausible for the \ha\ feature.
The inferred spot at the back of the disc is a consequence of the same
absorption component, found weakly in \ha, which depletes the broad
emission component, leaving a residue which maps to the extended spot in
Fig.~6. 

Before~back-projecting~the suitably-filtered \heired\ data, the interstellar 
features contaminating the red wing of the line were interpolated over.
The resulting map, presented in Fig.\ 6, shows the features interpreted in 
the data as the secondary star component, the strong
absorption core and the broad emission residual. Since the individual 
pixels within the map are time-independent, the narrow secondary star
emission and the absorption core have not been reconstructed accurately. 
Hence, as is common, for nova-like emission, the computed profiles of Fig.\ 7
do not provide a good representation of flux variations in the data.

If incorporated into the model of Keplerian disc and ballistic accretion
stream, both the map and computed profiles indicate that the absorption
occurs on the trailing side of the disc, although it does not appear to
coincide with either the gas stream trajectory or the Keplerian trajectory
off the stream. Although there is some freedom to arrange the conjunction
ephemeris and binary parameters according to the irradiation distribution
on the inner face of the secondary star, there are no plausible
configurations to allow the stream trajectories to coincide with the
absorption velocities. However, absorption strength appears 
correlated with the continuum hump displayed by the optical continuum (G94 and
Fig.\ 3), suggesting that the feature could be related to the
bright spot or accretion stream. A possible interpretation of its velocity
is that the disc is thicker downstream from the bright spot than upstream,
and the disc rim at these azimuths is able to eclipse the inner disc --
consistent with the model suggesting that it is the bright spot which is
being eclipsed by the secondary star. The absence of absorption at the likely
bright spot velocity is perhaps a consequence of core {\it filling} by
emission from the spot or stream.  The data is not of the desired
quality to favour or refute this idea; more precise light curves of the
lines features during eclipse are required for this purpose. 

Certainly, the line distribution does not provide a unique emissivity
model, however, the strength of the above interpretation is that it
considers only accretion elements that are well-founded in both theory and
observation. 

\section{Discussion}

The present data can be described by a model where the bright spot is the 
major contributor driving orbital phenomena in the nova-like variable
\pg. The bright spot appears
to be eclipsed by the secondary star, and low-excitation lines are
dominated by a variable component, perhaps from the spot region, which 
occults the
inner disc and provides a confusing set of emission profiles to
deconvolve.  Fig.\ 1 illustrates that there are a number of 
outlying eclipse times from the 
best linear orbital ephemeris. It is not implausible that these outliers are
the result of misleading statistics. However, it may be that this distribution
is not best fit by a linear ephemeris and is a result of variable
mass transfer from the secondary star (Robinson, Shetrone 
\& Africano 1991; Applegate 1992). If the emission model suggested by this 
data proves to be accurate, long-term photometric and
spectroscopic monitoring of bright spot intensities and relative disc sizes,
provided by eclipse times, may provide an opportunity
to study the accretion flows response to variations in mass transfer rate
from the secondary star. 
Further eclipse times are required to improve the statistics of the $O-C$ 
diagram before the significance of the linear ephemeris may be suitably tested.

\section{Conclusions}

The UX~UMa nova-like classification made by G94 is consistent with the
observation made in this paper that \pg\ is a lower inclination analogue
of UX~UMa (Kaitchuck, Honeycutt \& Schlegel 1983) and RW~Tri (Still
et~al.\ 1995). Lines are interpreted as composites of the bright
spot and inner accretion disc in absorption, and the irradiated inner face
of the secondary star and outer disc, or a disc wind, in emission.  If
unrecognised, the bright spot component, provides a confusing emission
distribution which is at odds with models of disc accretion. Provided the
identification of secondary star emission is correct, a lower limit to the 
radial velocity of the donor star is 280\,\kmsec, and the eclipse found in the
optical continuum must be dominated by the bright spot rather than the
disc. Further eclipse times are required to determine the significance of
outliers from the best linear orbital ephemeris, which may be indicative of
variable mass transfer from the secondary star.

\subsection*{Acknowledgments}

We thank Tom Marsh for the use and support of his spectral reduction
packages {\sc pamela}, {\sc molly} and {\sc doppler}, and Keith Horne and
Paul Bennie for comments and suggestions. The Isaac Newton Group of
telescopes are operated on the island of La Palma by the Royal Greenwich
Observatory in the Spanish Observatorio del Roque de los Muchachos of the
Instituto de Astrof\'{\i}sica de Canarias.

\bsp

\end{document}